\newcommand{\figref}[1]{Fig.~\ref{#1}}
\newcommand{\eqnref}[1]{Eq.~(\ref{#1})}

\documentclass[pra,aps,reprint,showpacs,twocolumn]{revtex4-1}
\usepackage{times}
\usepackage{amssymb}
\usepackage{amsmath}
\usepackage{mathrsfs}
\usepackage{graphicx}
\usepackage{dcolumn}
\usepackage{color}
\usepackage{bm}

\begin{document}

%******************************************************************************************************************************************************************************
%*****************************************************************************************************************************************************************************
\title{Enhancing quantum transduction via long-range waveguide mediated interactions between quantum emitters}
\author{Vincent E. Elfving*, Sumanta Das, and Anders S. S\o rensen}

\affiliation{Center for Hybrid Quantum Networks (Hy-Q), Niels Bohr Institute, University of Copenhagen, Blegdamsvej 17, 2100 Copenhagen \O, Denmark}
\date{\today}

%******************************************************************************************************************************************************************************
%*****************************************************************************************************************************************************************************
\begin{abstract}
Efficient transduction of electromagnetic signals between different frequency scales is an essential ingredient for modern communication technologies as well as for the emergent field of quantum information processing. Recent advances in waveguide photonics have enabled a breakthrough in light-matter coupling,  where individual two-level emitters are strongly coupled to individual photons. Here we propose a scheme which exploits this coupling to boost the performance of transducers between low-frequency signals and optical fields operating at the level of individual photons. Specifically, we demonstrate how to engineer the interaction between quantum dots in waveguides to enable efficient transduction of electric fields coupled to quantum dots. Owing to the scalability and integrability of the solid-state platform, our transducer can potentially become a key building block of a quantum internet node. To demonstrate this, we show how it can be used as a coherent quantum interface between optical photons and a two-level system like a superconducting qubit.
\end{abstract}

\maketitle

%******************************************************************************************************************************************************************************
%*****************************************************************************************************************************************************************************
Transduction of information between physical systems operating at different energy scales is of immense technological importance. In particular, efficient transduction between the microwave and optical domains is an essential requirement for telecommunication networks. The advent of quantum communication technologies necessitates analogous transduction devices capable of coherent information transfer for quantum fields. Possible applications of such devices range from a quantum internet \cite{oriRepeater,teleportation,crypto,Kim08, NV2} and distributed quantum computing \cite{distributed1,distributed2}, to quantum metrology \cite{sensing1,sensing2,metrology}.

Due to the large range of applications, several different methods for implementing quantum transducers have been investigated. A large class of these rely on the use of nanomechanical systems \cite{mech1,mech2,mech3,mech4,mech5} or direct electro-optical coupling \cite{MWtoPhot3,MWtoPhot5}.  Other proposals exploit  quantum emitters with both microwave and optical transitions \cite{MWtoPhot1,MWtoPhot2,MOProp1,NV1,NV2,NV3}. Many of these rely on weak magnetic interactions to single emitters, but exploit strong coupling to ensembles of emitters (see Refs. \cite{MOProp1,zollerPolar} for exceptions). Despite these efforts, efficient low noise quantum transduction of low-frequency excitations to the optical regime remains elusive.

Common to the approaches discussed above is the use of strong optical control fields. This can be a major source of light-induced decoherence \cite{decoherenceQuasi}, since even the absorption of a single optical photon can have a significant influence on a quantum system operating in the microwave regime. Furthermore strong light fields pose a filtering problem since weak quantum fields have to be distinguished from strong background signals.  It is therefore highly desirable to work at very low light levels. 

In this Rapid Communication, we propose an electrically coupled quantum transducer that works at the few-photon level. The principal elements of our transducer are semiconductor quantum dots (QDs) grown in photonic crystal waveguides with transform-limited linewidth \cite{transformLimited} and experimentally demonstrated coupling efficiency $\beta$ up to 98\% \cite{lodahl}. We show that such high coupling efficiencies enable a high transduction efficiency due to the strongly suppressed loss rate out of the waveguide. A key feature of our transducer is that it works efficiently even when all light fields contain at most a few photons. We achieve this by engineering the waveguide mediated interactions between multiple QDs. This low light level minimizes the light-induced decoherence of the quantum systems, as well as ensures  negligible background light in the transduced signal. 

\begin{figure}[ht]
	
	\centering
	\includegraphics[width=0.87\linewidth]{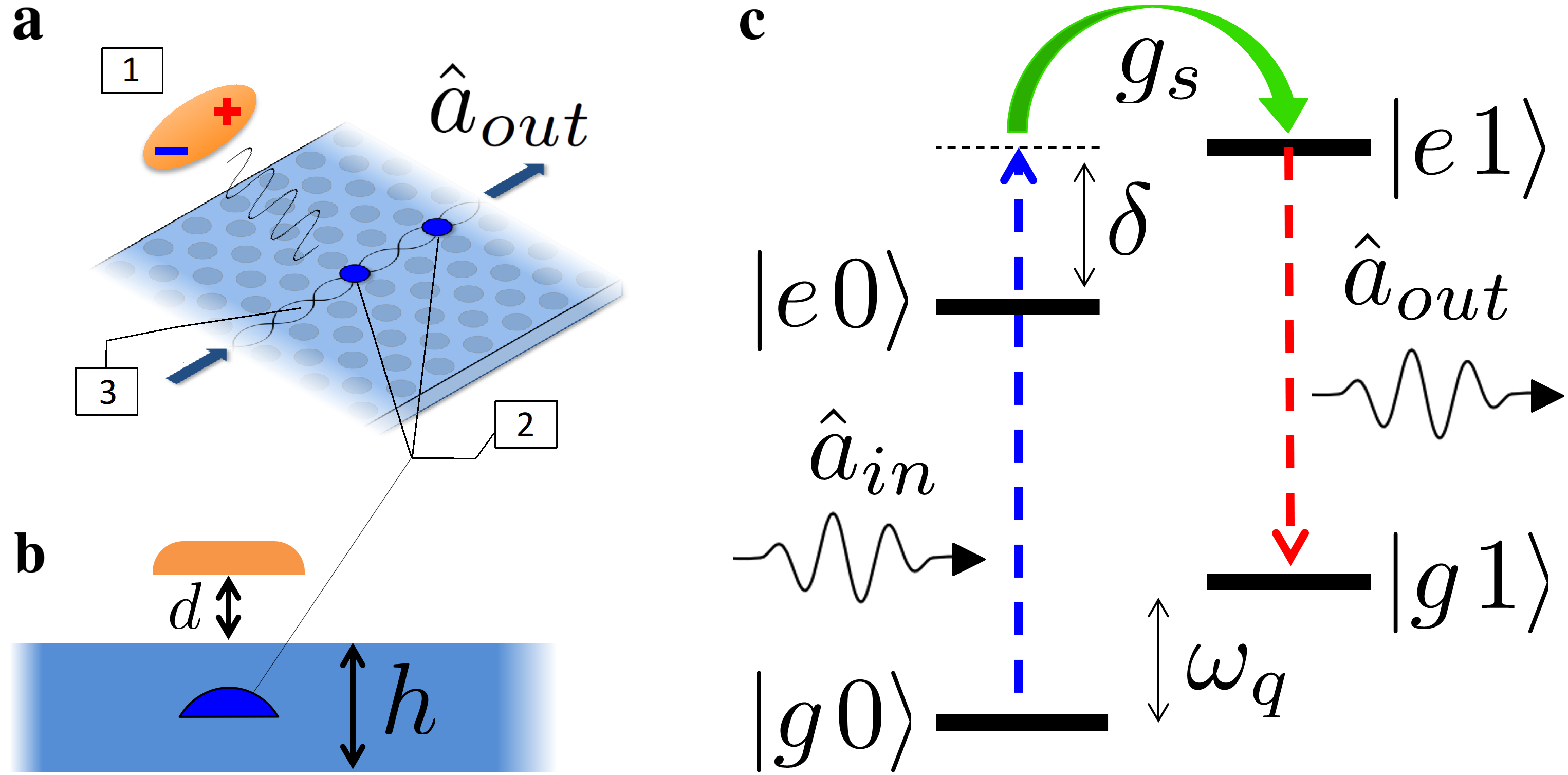}
	\caption{\textbf{a} A two-level system (TLS) represented by an oscillating electric dipole (1) is electrically coupled to a semiconductor QD (2) inside a waveguide (3). We consider several configurations with varying number and locations of the QDs. \textbf{b} Side view of the transducer. The TLS (orange) is a distance $d$ away from the photonic crystal of thickness $h=140$nm, with a QD in the center (roughly corresponding to Ref. \cite{lodahl}). \textbf{c} Energy level diagram of a single-QD transducer, with $|e\rangle/|g\rangle$ being the QD excited/ground state respectively. The electrical coupling $g_s$ enables a Raman transfer between internal states $|0\rangle$ and $|1\rangle$ of the coherent TLS, separated by a frequency difference $\omega_q$, through the path indicated by arrows.}
	\label{schematic}
\end{figure}

As a particular application we show how to exploit the proposed transducer as a quantum interface between optical photons and superconducting qubits. Related approaches were recently proposed using  two nearby dipole-coupled molecules \cite{DasPRL17} or a double QD molecule \cite{imamoglu}. As opposed to these systems, experimental demonstrations of QD-waveguide interfaces have shown more efficient coherent coupling to traveling light fields \cite{lodahl}. This strongly enhances the transduction efficiency in our scheme. We also show that long-range waveguide mediated interactions allows engineering super- and sub-radiant states \cite{Dicke,haroche,kaiser,Solano17,goban,laurat2} between distant QDs, which can be exploited to improve the transduction without requiring engineering complicated near-field interactions.

In \figref{schematic}a, we show schematically our proposed quantum transducer. It consists of three components; a 1D waveguide for efficient confinement of the optical mode; one or more QDs coupled to the photonic mode with high efficiency; and finally a nearby oscillating electric dipole, which electrically couples to a QD excitation. For specificity, we focus on transduction from a coherent two level system (TLS) with a dipole allowed transition at a non-optical frequency, e.g. in the GHz regime. Examples of such coherent TLSs include superconducting qubits \cite{coherencecpb,nakamura,devoretReview} and singlet-triplet states in double-QD structures \cite{singtrip1,singtrip2,singtrip3}.

To begin with we consider a transducer with a single QD in the photonic waveguide coupled electrically to the oscillating dipole of a coherent TLS (\figref{schematic}a). The total Hamiltonian $\mathcal{H}$ describing this system can be written $\mathcal{H}=\mathcal{H}_0+\mathcal{V}_1+\mathcal{V}_2$ where $\mathcal{H}_{0}=\omega_{d}\hat{\sigma}^\dagger \hat{\sigma} + \sum_k \omega_{k} \hat{a}_k^\dagger \hat{a}_k +\omega_q \hat{\eta}_z$ ($\hbar=1$) with $\omega_q$, $\omega_{d}$ and $\omega_{k}$ being the transition frequency of the TLS, QD and photonic modes, respectively. $\mathcal{V}_1=\sum_k g_{k} (\hat{\sigma} \hat{a}_k^\dagger+ \hat{\sigma}^\dagger \hat{a}_k)$ represents the interaction between the QD and optical fields, where $g_{k}$ is the coupling of the 2-level QD to the $k$'th mode with annihilation operator $\hat{a}_k$, and $\hat{\sigma}=|g\rangle\langle e|$ is the standard lowering operator of the QD. The TLS is represented by the Pauli-X and Z operators, where $\hat{\eta}_x=|1\rangle\langle 0|+|0\rangle\langle 1|$ and $\hat{\eta}_z=(|1\rangle\langle 1|-|0\rangle\langle 0|)/2$ with $|0\rangle$ and $|1\rangle$ being the internal states. As we assume that the TLS has a dipole allowed transition, there will be an associated electric field of the form $\hat{E}= \vec{E}(r) \eta_x$. QDs are known to exhibit sizable Stark shifts of their excited levels, corresponding to a dipole moment up to $|\vec{p}|\approx0.4\text{ e}\cdot\text{nm}$ \cite{stark}, for an In(Ga)As QD. The proximity to the TLS thus leads to an interaction of the form $\mathcal{V}_2 = g_s \hat{\eta}_x \hat{\sigma}^\dagger \hat{\sigma}$ with $g_s\equiv\vec{p}\cdot \vec{E}/\hbar$. As we discuss below this interaction can be sizable. We estimate $g_s=2\pi \times( 0.4-1) \text{ GHz}$ for a Cooper pair box (CPB) \cite{coherencecpb} and expect similar results for double-QD structures \cite{singtrip1,singtrip2,singtrip3}, since the oscillating charge is similar in this case. For typical QDs this coupling is larger than their total decay rate $\Gamma\approx 2\pi \times 150$ MHz. The system is thus in a strong coupling regime $g_s>\Gamma$ enabling an efficient transducer design.
 
We consider a Raman transition between the states of the combined TLS-QD system (\figref{schematic}c) via a single incoming optical photon in the waveguide with central frequency $\omega_p$. The Raman transition entangles the frequency of a scattered weak photon pulse with the internal state of the TLS, thereby achieving coherent transduction between the two systems. To study the dynamics of the transducer, we apply the formalism of Ref. \cite{Das17,Reiter12} and find that the total scattered field is given by $\hat{a}_{out}=\left[1+i\hat{\sigma}_{00} \mathcal{S}_{00}+i\hat{\sigma}_{01} \mathcal{S}_{10}\right] \hat{a}_{in}+\mathcal{F}$, 
where $\hat{a}_{in}$ and $\hat{a}_{out}$ are the right-going input and output field operators respectively, $\hat{\sigma}_{01}\equiv|1 \rangle\langle 0|$ and $\hat{\sigma}_{00}\equiv|0 \rangle\langle 0|$ are respectively the Heisenberg operators for the coherence and population of the ground state of the combined QD-TLS system. $\mathcal{F}$ is a noise operator and $\mathcal{S}_{00}$ and $\mathcal{S}_{10}$ are the scattering amplitudes for the transitions from state $|0\rangle$ to  $|0\rangle$ and  $|1\rangle$, respectively, and are calculated from the total contribution of all excited states. 

We assume that the ground states are sufficiently separated in energy compared to the width of the incoming photon pulse, so that scattered photons can be filtered spectrally and the only contribution to `red' photons comes from the term $i\hat{\sigma}_{01} \mathcal{S}_{10} \hat{a}_{in}+\mathcal{F}$. Detection of such a red-detuned photon heralds a flip of the TLS; the Raman scattering detection probability $P_R$ for a single-photon input can be found by the expectation value of the photon-number operator of the red field 
$P_R=\int_{0}^{T}\langle \hat{a}_{out}^\dagger\hat{a}_{out}\rangle_{\text{red}}\hspace{2pt}dt=|\mathcal{S}_{10}|^2$ where we have normalized the incoming pulse of duration $T$ to contain a single photon. Note that the quantum vacuum noise term $\mathcal{F}$ in $\hat{a}_{out}$ does not contribute to the photon number expectation value. The scattered right-going mode (transmission) is equal in magnitude to the scattered left-going mode (reflection). Similarly the QD couples equally to left and right propagating modes. Experimentally, the scattered red field amplitude can therefore be twice enhanced by combining both modes on a beamsplitter \cite{singlephotonChang}, or using a single-sided waveguide \cite{singlesided}, corresponding to a factor of 4 improvement in the success probability.

From the scattering formalism of Ref. \cite{Reiter12}, we find the scattering matrix element $\mathcal{S}_{10}=\frac{\Gamma_{1D}}{2} \langle e1| \hat{\mathcal{H}}_{nh}^{-1} |e0\rangle$ and therefore $P_R=\Gamma_{1D}^2|\langle e\hspace{1pt} 1|\mathcal{H}_{nh}^{-1}|e\hspace{1pt}0\rangle|^2$. Here, the scattering dynamics of the excited subspace of the system are fully absorbed into an effective non-Hermitian Hamiltonian ${\mathcal{H}}_{nh}={\mathcal{H}}_{e}-\frac{i}{2}\sum_j \mathcal{L}_j^\dagger \mathcal{L}_j$, where $\mathcal{H}_{e}$ describes the energies and couplings in the excited subspace of the total Hamiltonian, and $\mathcal{L}_j$ are the Lindblad operators associated with interactions with the environment. For our single-QD system, $\mathcal{H}_e=-\delta \hat{\mathbb{I}}+\omega_q (\hat{\mathbb{I}}-\hat{\eta}_z)/2 +g_s \hat{\eta}_x$, with $\delta=\omega_{d}-\omega_p$ being the photon-QD detuning and $\hat{\mathbb{I}}\equiv |e0\rangle \langle e0| + |e1\rangle \langle e1|$. The operators representing the decay dynamics of the system are defined as $\mathcal{L}_k=\sqrt{\Gamma_k} |g\rangle \langle e|$ with $\Gamma_k$ being the QD decay into and out of the waveguide with rates $\Gamma_{1D}$ and $\gamma'$ respectively.

We maximize $P_R$ as a function of the detuning between input field and QD, yielding the resonance conditions $\delta_\pm=(\omega_q\pm\sqrt{\omega_q^2+g_s^2-\Gamma^2})/2$. At these resonances, $P_R^{(1QD)}=\beta^2 g_s^2/(g_s^2+\omega_q^2)$, and in the strong/weak coupling limits we find
\begin{eqnarray}
P_R^{(1QD)} &\approx&  \beta^2 \hspace{23pt} \text{for} \hspace{10pt} g_s\gg\omega_q, \label{PR1qda}\\
P_R^{(1QD)} &\approx&  \beta^2\frac{g_s^2}{\omega_q^2} \hspace{10pt} \text{for} \hspace{10pt} g_s\ll\omega_q, \label{PR1qdb}
\end{eqnarray}
where $\beta=\Gamma_{1D}/(\Gamma_{1D}+\gamma')$ describes the QD-waveguide coupling efficiency. \eqnref{PR1qda} expresses the striking advantage that can be obtained by exploiting strong coupling of emitters with a waveguide. For $\beta$ approaching unity, a transducer can be constructed where even a single optical photon is sufficient to efficiently transduce a low frequency signal. This minimizes any possible decoherence induced by light.

Many of the qubit systems relevant for this transduction scheme operate in the microwave (GHz) regime, which is larger than the maximal estimated coupling $g_s\lesssim 2\pi \times 1$ GHz. The resulting reduction of $P_R$ in \eqnref{PR1qdb} arises because the electric dipole moment of the TLS is linked to a transition between two energy levels. The QD thus feels an oscillating field, and this averages out the coupling. To counter this effect, we propose to engineer the excited subspace using multiple coupled emitters. This exploits the high $\beta$-factor achievable for QDs in photonic crystal waveguides to get strong waveguide mediated interactions between distant QDs. With two QDs we show how to engineer an increased interaction time, thereby improving the effective coupling. By using four QDs one can even engineer a resonant Raman transition, thereby avoiding the averaging effect. 

\begin{figure}[ht]
	\centering
	\includegraphics[width=1\linewidth]{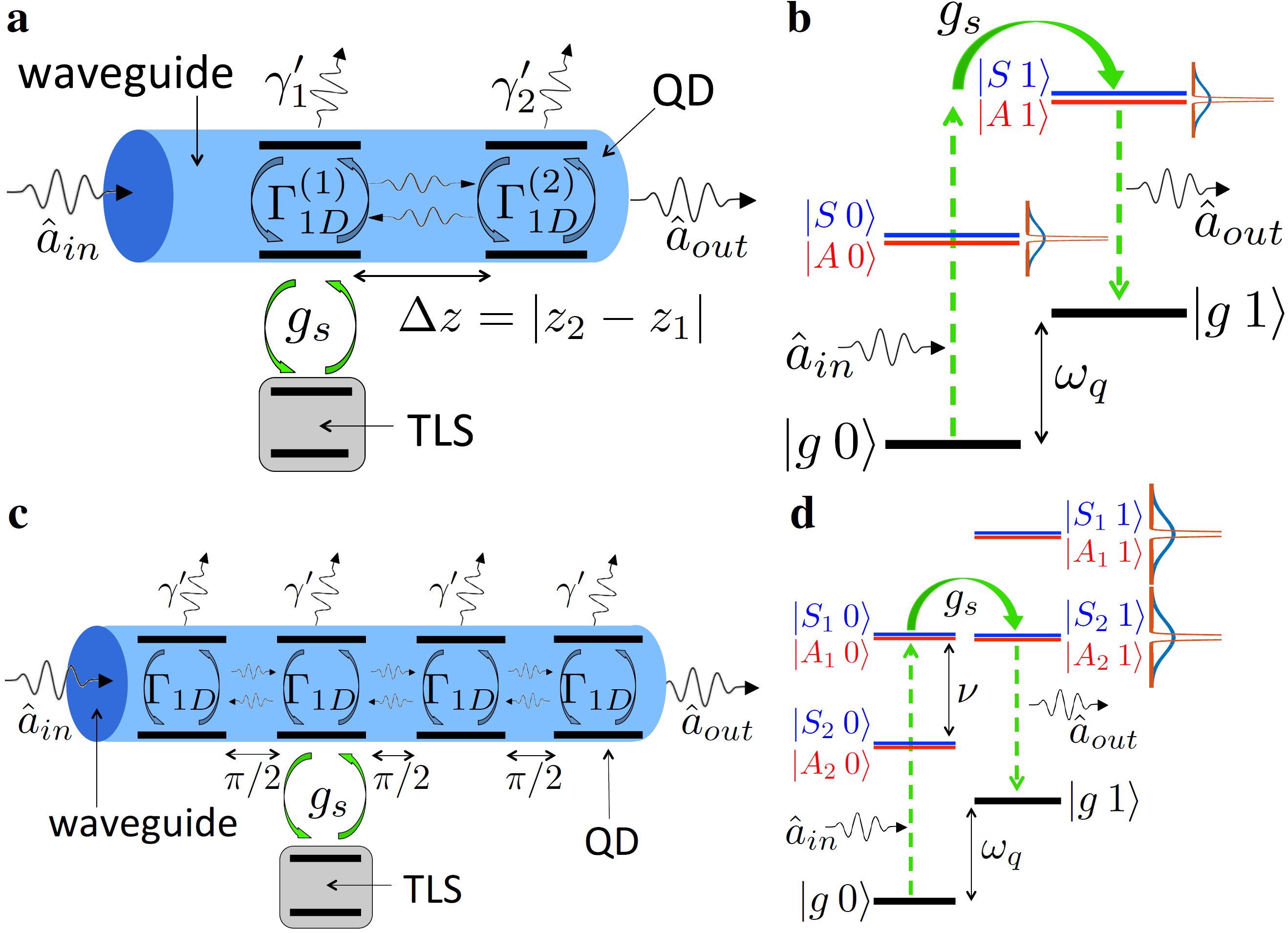}
	\caption{\textbf{a} Transducer based on two QDs in a waveguide. One of the QDs is electrically coupled to the TLS. \textbf{b} Energy level diagram of the combined QD-TLS system. $\{|0\rangle,|1\rangle\}$ represent the TLS internal states. $\{|A\rangle,|S\rangle\}$ represent the (anti-)symmetric states of the two-QD coupled system. The incoming photon is tuned into resonance with $|A\hspace{2pt}1\rangle$. \textbf{c} Four QDs in a waveguide are spaced equidistantly such that $k\Delta z=\pi/2$. One of the central QDs is coupled to a coherent two-level system. \textbf{d} Energy level diagram of the combined QD-TLS system for 4 QDs, with two sets of (anti-)symmetric eigenstates spaced by $\nu\approx1.3\Gamma_{1D}$, decaying at enhanced(reduced) decay rates.}
	\label{interferenceScheme}
\end{figure}

We first consider two emitters in a 1D waveguide (see \figref{interferenceScheme}a). The photonic field in the waveguide induces long-range interactions between the two. In Ref. \cite{Das17} this interaction is included as part of a non-Hermitian Hamiltonian of the single-excitation subspace for the bare two-emitter system $\mathcal{H}_{nh}=-\delta \hat{\mathbb{I}}+(\Delta/2-i\Gamma_1/2) |eg\rangle\langle eg|+(-\Delta/2-i\Gamma_2/2) |ge\rangle\langle ge|+\Omega (|eg\rangle\langle ge|+|ge\rangle\langle eg|)$, where $|eg\rangle\equiv|e\rangle_1\otimes|g\rangle_2$, $\hat{\mathbb{I}}\equiv |eg\rangle\langle eg|+ |ge\rangle\langle ge|$, $\Delta$ is the difference between the emitters' transition frequencies, and $\delta$ is the detuning between the incoming photon and mean QD transition frequency. $\Gamma_i=\Gamma_{i,1D}+\gamma_i'$ is the decay rate of emitter $i$, which consists of the decay rate into the waveguide $\Gamma_{i,1D}$ and to the environment  $\gamma_i'$. The collective complex coupling term $\Omega=-i\sqrt{\Gamma_{1,1D}\Gamma_{2,1D}}\exp [i k\Delta z]$ is the  waveguide-mediated coherent coupling between emitters and collective decay. $\Delta z=|z_2-z_1|$, and $z_i$ is the position of emitter $i$ \cite{Can15}. 

We diagonalize the Hermitian part of the Hamiltonian $\mathcal{H}_{nh}$ and find (anti-)symmetric eigenstates ($|A\rangle=\xi_1|eg\rangle-\xi_2|ge\rangle$) $|S\rangle=\xi_2|eg\rangle+\xi_1|ge\rangle$. Here coefficients $\xi_{1,2}$ depend on the collective coupling $\Omega$ and detuning $\Delta$ between the emitters. The decay rates $\Gamma_j=\gamma_j'+\Gamma_{j,1D}$ with $j=A,S$ consist of a part going outside the wave guide $\gamma_A'=\xi_1^2 \gamma_1'+\xi_2^2\gamma_2'$, $\gamma_S'=\xi_2^2 \gamma_1'+\xi_1^2\gamma_2'$ and a part going into the wave guide $\Gamma_{A,1D}=\xi_1^2 \Gamma_{1,1D}+\xi_2^2\Gamma_{2,1D}-2\xi_1\xi_2\sqrt{\Gamma_{1,1D}\Gamma_{2,1D}}\cos (k\Delta z)$, $\Gamma_S=\xi_2^2 \Gamma_{1,1D}+\xi_1^2\Gamma_{2,1D}+2\xi_1\xi_2\sqrt{\Gamma_{1,1D}\Gamma_{2,1D}}\cos (k\Delta z)$.  For $\cos (k\Delta z)\approx 1$ the decay into the waveguide can show almost complete destructive interference $\Gamma_{A,1D}\approx 0$ resulting in the anti-symmetric state having a much longer lifetime, $\Gamma_S\gg\Gamma_A$, if  the decay is dominated by the decay into the waveguide $\Gamma_{1,1D},\Gamma_{2,1D}\gg\gamma_{1}',\gamma_{2}'$. By adjusting the relative detuning $\Delta$ to ensure $\xi_1^2\Gamma_{1,1D}=\xi_2^2\Gamma_{2,1D}$, this increase in lifetime is achievable even if the emitters' decay rates differ.

In the following we assume that the TLS only couples to a single QD, while the other QD is placed far away and does not directly influence the TLS. This adds $\mathcal{H}_e=\omega_q (1-\hat{\eta}_z)/2 +g_s \hat{\eta}_x \otimes|eg\rangle\langle eg|$ to the Hamiltonian, similar to the single-QD-qubit system. As described above, the excited states of the two QDs couple and hybridize into (anti-)symmetric eigenstates (\figref{interferenceScheme}(b)). The transition pathway  consists of 4 paths which contribute to the output field amplitude. These contributions can be conveniently summed using the formalism of Ref. \cite{Das17}. 

If we tune the incoming photon to be in resonance with the antisymmetric state $|A\hspace{2pt}1\rangle$, the Raman transition rate will be dominated by a single path, $|g\hspace{2pt}0\rangle\rightarrow|S\hspace{2pt}0\rangle\rightarrow|A\hspace{2pt}1\rangle\rightarrow|g\hspace{2pt}1\rangle$. The probability for this path can be written in the form $P_R\approx\Gamma_{A,1D}\Gamma_{S,1D} |\langle A1|\mathcal{H}_{nh}^{-1}|S0\rangle|^2$, using similar techniques as described above for the single-QD case. For $g_s\ll\omega_q$ we find the resonance condition 
$\delta=\omega_q - \sqrt{\Gamma_{1,1D} \Gamma_{2,1D}}\sin (k\Delta z)/2+g_s^2/(4\omega_q)$, where
\begin{equation}\label{PR2sym}
P_R^{(2QD)}\approx\left(\frac{\Gamma_{S,1D}}{\Gamma_A}\right)\left(\frac{\Gamma_{A,1D}}{\Gamma_{A}}\right)\left(\frac{g_s^2}{\omega_q^2}\right).
\end{equation}
Due to the first factor  $\Gamma_{S,1D}/\Gamma_A$ the Raman probability is increased by making  $|A\rangle$ longer-lived so that $\Gamma_{S,1D}\gg \Gamma_A$, which effectively increases the interaction time with the TLS. Because of the second factor $\Gamma_{A,1D}/\Gamma_A$, there exists an optimal value $\Gamma_{A,1D}\approx \Gamma_A/2$. Here we find $P_R^{(2QD)}\approx \Gamma_{S,1D}/2\Gamma_A\cdot g_s^2/\omega_q^2$ showing a strong enhancement compared to \eqnref{PR1qdb} if we can generate a large difference in the super- and subradiant decay rates $\Gamma_S\gg\Gamma_A$. 

\begin{figure}[ht]
	\centering
	%	\begin{tabular}{l l}
	%		\textbf{a} & \textbf{b}\\
	%	\includegraphics[width=0.4\linewidth]{Fig3.pdf} & \includegraphics[width=0.6\linewidth]{Fig4.pdf}
	%	\end{tabular}
	\includegraphics[width=1\linewidth]{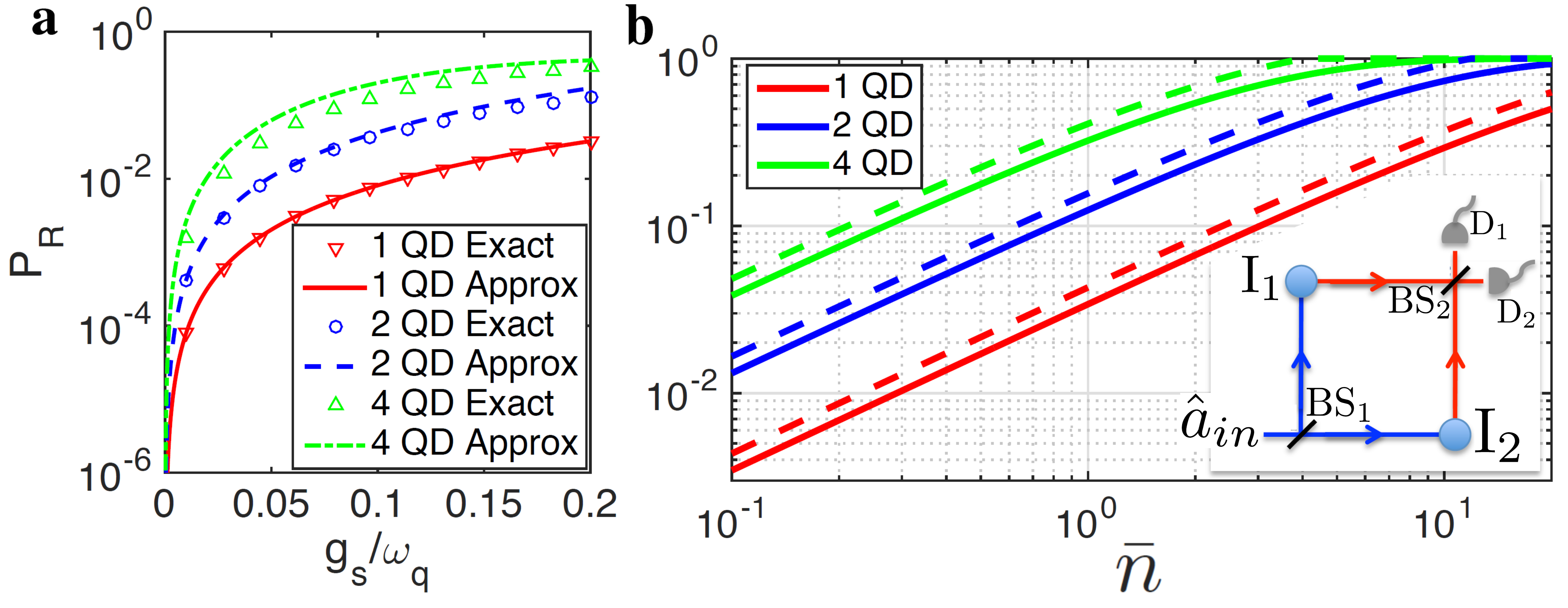}
	\caption{\textbf{a} Raman transition probability as a function of coupling strength $g_s$. The exact forms (markers) include all scattering pathways, while approximate forms (lines) include only the most significant pathway, as given by Eqs. \eqref{PR1qdb}, \eqref{PRinterfere2QD} and \eqref{PR4qd}. For the 1-QD (red) and 2-QD (blue) cases, we set $\omega_q=2\pi\times5$ GHz, $\Gamma_{1D}=1 \text{ ns}^{-1}$, $\beta=0.9$. In the 4-QD case (green), we assume a Purcell enhanced emitter decay rate $1/\Gamma_{1D}=1.27/\omega_q\approx250$ ps and $\beta=0.9$. We optimize the detunings $\delta$ and $\Delta$ in the full numerics and fix the inter-QD spacing $k\Delta z=n\pi$ (with integer n) in the 2-QD case and to $k\Delta z=\pi/2$ in the 4-QD case. \textbf{b} Infidelity ($1-F$, solid lines)  and success probability $P_{\text{suc}}$ (dashed lines) of entanglement generation for a coherent pulse input, as a function of average photon number. Detection efficiency $\eta=0.7$, coupling $g_s=2\pi\times1$ GHz, and the other parameters as in \textbf{a}.}\label{124qd}
\end{figure}

Assuming for concreteness that the emitters are identical, the optimum is reached for $\Gamma_{A,1D}=\beta(1-\beta)\Gamma$,  which can be met for any emitter spacing fulfilling $\cos(k\Delta z)\geq \beta$ by adjusting $\Delta$, e.g. using an external field (note that deterministic placement of a QD in a waveguide with a precision less than 10 nm has been achieved \cite{tommaso}). Here $\Gamma=\Gamma_1=\Gamma_2$ and $\beta$ is defined analogous to the one-emitter case. For this condition, we find 
\begin{equation}\label{PRinterfere2QD}
P_R^{(2QD),opt}\approx\frac{\beta^2}{1-\beta^2}\Big(\frac{g_s}{\omega_q}\Big)^2.
\end{equation} 
Comparing \eqnref{PRinterfere2QD} to \eqnref{PR1qdb} there is a factor $1/(1-\beta^2)$ improvement in transduction efficiency compared to the single QD case. For $\beta=0.9$, this is a factor 5 improvement. For $\beta=0.98$, as experimentally demonstrated in Ref. \cite{lodahl}, this is a factor 25. This conclusion is verified in \figref{124qd}a, where we compare \eqnref{PRinterfere2QD} to the single QD case and the full calculation including all paths. 

For non-identical emitters we have verified numerically that there is an enhancement of the Raman probability as discussed below \eqnref{PR2sym}. In the supplementary information \cite{supp} we show that for two emitters with identical $\beta$ but different rates $\Gamma_i$, \eqnref{PRinterfere2QD} remains an excellent approximation regardless. 

The main limitation of the above scheme is that the relevant transition is still far off-resonant, resulting in the factor $(g_s/\omega_q)^2$ suppressing the efficiency. With \textit{four} QDs we can further engineer the energy of the long-lived states. By tuning the frequency of two such states in resonance to the TLS energy $\omega_q$ we can achieve an enhanced Raman probability.

To this end, we consider 4 QDs placed in a 1D waveguide (see \figref{interferenceScheme}c). At zero mutual detuning between the emitters and assuming equal decay rates we have identified an optimum at $k\Delta z=\pi/2$ between each QD. There, we find two bright and two dark states, with an energy splitting $\nu=\sqrt{\frac{1}{2}(\sqrt{5}+1)}\Gamma_{1D}\approx 1.27\hspace{3pt}\Gamma_{1D}$ between them. The dark states exhibit suppressed decay rates $
\Gamma_{1,1D}=\Gamma_{2,1D}=(1-\alpha_{4})\Gamma_{1D}$ into the 1D waveguide with $\alpha_4=\sqrt{\frac{1}{2}(\sqrt{5}-1)}\approx0.79$. The resonance condition in the excited state manifold for a Raman process $|0\rangle\rightarrow|1\rangle$ is achieved for $\omega_q=\nu=1.27\Gamma_{1D}$ (\figref{interferenceScheme}d). This condition can be met either by choosing a TLS with matching transition energy and/or Purcell enhancing the waveguide decay rate \cite{QDpurcell1,QDpurcell2,Solano_arXiv}.

The resulting Raman probability (\figref{interferenceScheme}d) has an optimum at $\delta=\omega_q$, where
\begin{equation}\label{PR4qd}
P_R^{(4QD)}\approx\frac{0.11(g_s/\omega_q)^2}{((g_s/\omega_q)^2+(0.79/\beta-0.62)^2)^2}.
\end{equation}
which we plot in \figref{124qd}a together with an exact expression involving all pathways, for all considered transducer configurations. In the limit of weak coupling $g_s\ll\omega_q$, we find $P_R\approx 101(g_s/\omega_q)^2$ for $\beta=0.98$, and $P_R\approx 27(g_s/\omega_q)^2$ for $\beta=0.9$. Compared to the single-QD interface at $\beta=0.98$, this is more than a two orders-of-magnitude improvement on the scaling with $(g_s/\omega_q)^2$; it represents a fourfold improvement over the two-QD case.

\begin{figure}[h]
	\includegraphics[width=1\linewidth]{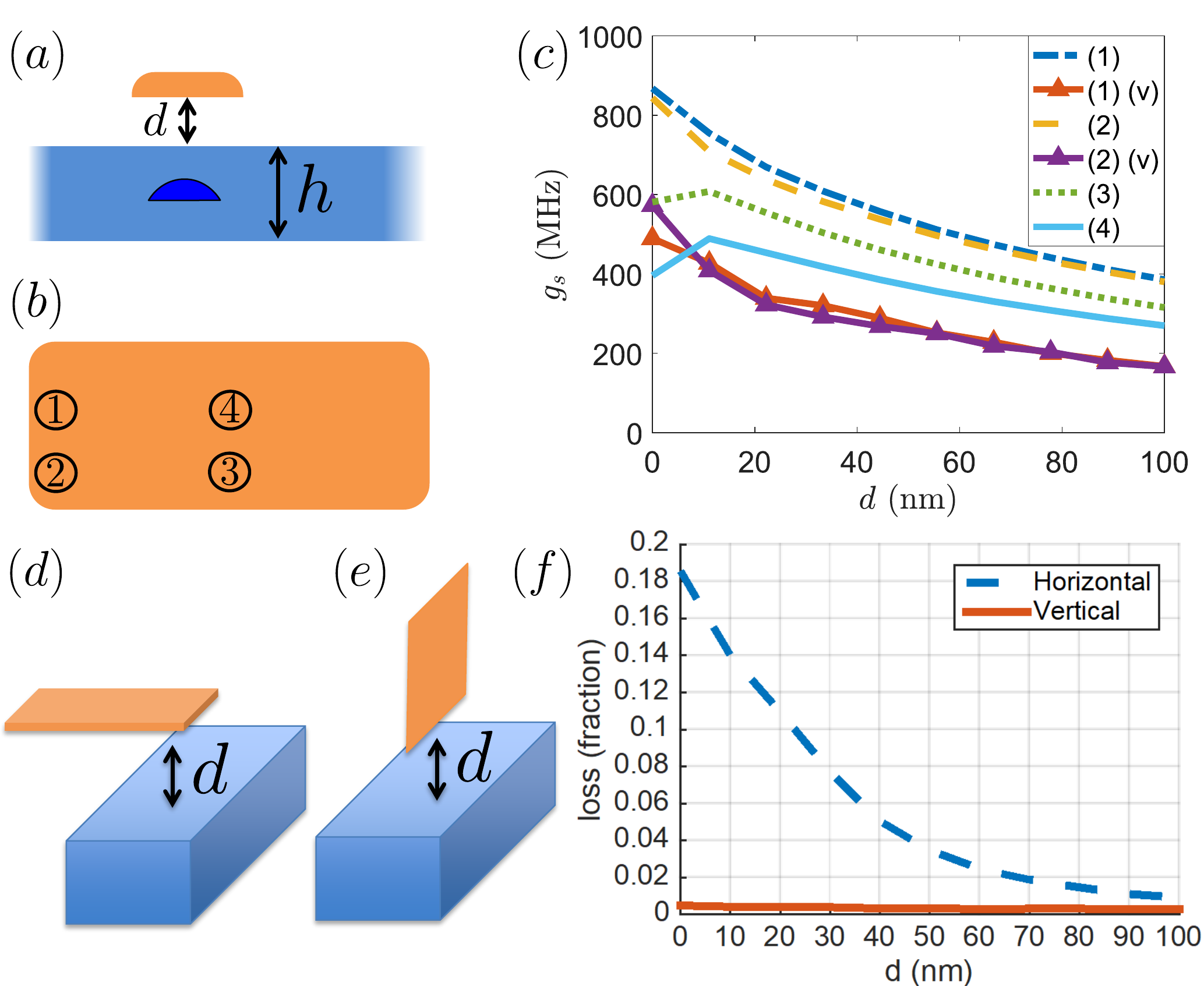}
	\caption{(a) The qubit (orange) is positioned at a distance $d$ away from the waveguide (blue) of thickness $h=140$ nm with a quantum dot in the middle. (b) Top view of the qubit, with numbers labeling the relative position of the QD underneath the qubit. Assumed dimension are: short edge $200$ nm, long edge $700$ nm, and thickness $20$ nm. (c) Coupling $g_s$ derived from an electrostatic simulation of the device. Different relative positions of the QD are labeled with indices defined in (b). Suffix (v) refers to the vertical geometry (e), whereas the remaining curves are for the horizontal geometry (d). (f)  Simulation of optical absorption as a function of distance $d$, for the two different configurations (d) and (e), assuming the short edge of the qubit to be aligned with the center of the waveguide.}
	\label{finalfigure}
\end{figure}

An important application of our proposed transducer is as a quantum interface between an optical photon and a microwave superconducting qubit.
Similar propsals have been considerd in Refs. \cite{DasPRL17,imamoglu}, but as discussed above the current approach takes advantage of the demonstrated high coupling efficiency in waveguides and avoids the need for engineering near field interactions. To get an estimate for the magnitude of the electric coupling $g_s$ between a SC qubit and a QD, we numerically simulate the electric field for a Cooper pair box (CPB) island of dimensions $700\times200\times20$ nm, similar to Ref. \cite{coherencecpb}, placed above a photonic crystal waveguide of height $h=140$ nm (\figref{finalfigure}a). The CPB qubit is defined by a single Cooper pair oscillating on and off a superconducting island. We simulate the electric field strength coming from two electrons on the island in order to estimate the Stark shift in the QD. Self-assembled In(Ga)As QDs with transform-limited linewidths \cite{transformLimited} and near-unity $\beta$ \cite{lodahl} have been reported to exhibit a Stark coefficient $|\vec{p}|=2\pi\times 100$ MHz/(kV/m) \cite{stark}. From the electric field strength we find a Stark coupling $g_s=2\pi\times0.4-1$ GHz for a separation of $d=0-100$ nm between qubit and waveguide, with the QD in the middle of the waveguide (\figref{finalfigure}c). 

We also simulate the optical field in this configuration with a 3D numerical integration of Maxwell's equations, and analyze the optical absorption by integrating the total power flow over the surface of the superconducting island. This yields the total absorbed fraction of the power, which directly translates into the photon absorption probability (\figref{finalfigure}f). We assume an optical wavelength of $\lambda=980$ nm (in air). We model the system as a 300 nm wide nano-beam waveguide made of GaAs with refractive index $n_{GaAs}=3.456$ \cite{gaas} suspended in vacuum. We assume negligible absorption in these media, while for the superconductor the real part of the refractive index is $n_{Al}=1.47$ and an extinction coefficient $\kappa=9.22$, corresponding to aluminum \cite{aluminum}. We find the absorption of the light-field into the qubit to be less than $1\%$ for the configurations with lower coupling strengths, and slightly higher for the largest coupling strengths. We emphasize that we consider rather simple  geometries, and carefully designed structures can likely improve these numbers. We note that, in the case of scattering between multiple QDs, the optical absorption is modified in a non-trivial way as  photons may bounce back and forth between QDs resulting in multiple passes. Detailed examination of this effect is beyond the scope here.

Including all scattering pathways in the calculation, for coupling in the low end of our estimate $g_s=2\pi\times400$ MHz, a CPB qubit with $\omega_q=2\pi\times5$ GHz, and $\beta=0.98$ we find $P_R\approx0.6$\%, $13$\% and $31$\% for the single-QD, 2-QD and 4-QD interfaces, respectively. For the strongest coupling of $g_s=2\pi\times1$ GHz we find $P_R\approx4$\%, $40$\%, $78$\% respectively.

The quantum transducer presents an ideal platform for long-distance entanglement. Using an interference protocol similar to Ref. \cite{MachZender}, we consider two transducers $I_1$ and $I_2$ placed in either arm of a Mach-Zehnder interferometer (inset in \figref{124qd}b). Photon scattering creates entanglement between the photon frequency and the SC qubit state. Mixing the red sideband fields on a BS and detecting a photon creates entanglement between the qubits. For a single-photon input, the protocol succeeds with probability $P_{\text{suc}}=\eta P_R$, where $\eta$ is the detection efficiency, and produces a maximally entangled state of fidelity $F=1$ once a photon is detected.

It is experimentally less challenging to use a weak coherent pulse with a low average photon number $\bar{n}$ instead. This reduces the fidelity, because the pulse may dephase or flip both qubits simultaneously. In \figref{124qd}b, we show the fidelity and success probability for a coherent input pulse, calculated using the approach of Ref. \cite{DasPRL17}. The considered Raman protocol for coherent inputs has an intrinsic requirement $1-F \geq 1-P_{\text{suc}}/\eta$. Our result is close to this limit, but has a slightly lower fidelity due to elastic (Rayleigh) scattering.  As shown in the figure, multi-QD transducers enable the generation of high quality entanglement for much lower mean photons numbers. Exploiting the waveguide mediated interactions thus reduces the possible detrimental decoherence of the SC qubit induced by the light, and allows for a near-deterministic interface between individual photons and SC qubits. The input pulse duration is mainly limited by the linewidth of the transitions and can be in the range of 50-100 ns, reducing the effect of decoherence. For comparison, superconducting qubits of the type considered here have demonstrated coherence times in the microsecond range \cite{coherencecpb}.

In summary we have shown that long-range waveguide mediated interactions can be exploited to boost the efficiency of quantum transducers.  As a direct application, the proposed device can be used to provide an on-chip interface between SC qubits and optical photons. This could facilitate a breakthrough in long-distance quantum communication via a quantum repeater network \cite{oriRepeater,teleportation,crypto,Kim08, NV2} and scaling of SC quantum computers by connecting them optically \cite{distributed1,distributed2}. Alternatively the proposed transducers can have applications for quantum limited sensing by exploiting efficient  optical detection of low frequency fields \cite{sensing1,sensing2,metrology}.
\begin{acknowledgments}
We gratefully acknowledge financial support from the European Union Seventh Framework Programme ERC Grant QIOS (Grant No. 306576), the Danish council for independent research (Natural Sciences), and the Danish National Research Foundation (Center of Excellence `Hy-Q', grant number DNRF139),
\end{acknowledgments}

%%%%%%%%%%%%%%%%%%%%%%%%%%%%%%%%%%%%%%%%%%%%%%%%%%%%%%%%%%%%%%%%%%%%%%%%%%%%%%%%%%%%%%%%%%%%%


\begin{thebibliography}{99}
	
	\bibitem{teleportation}
	S. Pirandola \textit{et al.}, Nat. Phot. {\bf 9}, 641 (2015).
	
	\bibitem{oriRepeater}
	H.J. Briegel,W. D\"{u}r, J.I. Cirac, and P. Zoller, Phys. Rev. Lett. {\bf 81}, 5932 (1998).
	
	\bibitem{crypto}
	H.K. Lo, M. Curty and Kiyoshi Tamaki, Nat. Phot. {\bf 8}, 595 (2014).
	
	\bibitem{NV2}
	E. Togan \textit{et al.}, Nature \textbf{466}, 730 (2010).
		
	\bibitem{Kim08}
	H.F. Kimble, Nature {\bf 453}, 1023 (2008).		
	
	\bibitem{distributed1}
	L.K. Grover, Preprint at https://arxiv.org/abs/quant-ph/9704012  (1997).
	
	\bibitem{distributed2}
	T. Pellizzari, Phys. Rev. Lett. \textbf{79}, 5242 (1997).
	
	\bibitem{sensing1}
	K. Zhang, F. Bariani, Y. Dong, W. Zhang, and P. Meystre, Phys. Rev. Lett. \textbf{114}, 113601 (2015)
	
	\bibitem{sensing2}
	J. A. Sedlacek \textit{et al.}, Nat. Phys. \textbf{8}, 819-824 (2012)	
	
	\bibitem{metrology}
	V. Giovannetti, S. Lloyd and L. Maccone, Nat. Phot. {\bf 5}, 222 (2011).
	
	\bibitem{mech1}
	K. Stannigel, P. Rabl, A. S. S{\o}rensen, P. Zoller, and M. D. Lukin, Phys. Rev. Lett. {\bf 105}, 220501 (2010).
	
	\bibitem{mech2}
	Sh. Barzanjeh, M. Abdi, G. J. Milburn, P. Tombesi, and D. Vitali, Phys. Rev. Lett. {\bf 109}, 130503 (2012).
	
	\bibitem{mech3}
	J. Bochmann, A. Vainsencher, D. D. Awschalom and A. N. Cleland, Nat. Phys. {\bf 9}, 712 (2013).
	
	\bibitem{mech4}
	R. W. Andrews \textit{et al.}, Nat. Phys. {\bf 10}, 321 (2014).
	
	\bibitem{mech5}
	T. Bagci \textit{et al.}, Nature {\bf 507}, 81 (2014).
	
	\bibitem{MWtoPhot3}
	A. Rueda \textit{et al.}, Optica {\bf 3}, 6 (2016).
	
	\bibitem{MWtoPhot5}
	M. Tsang, Phys. Rev. A {\bf 84}, 043845 (2011).
	
	\bibitem{MWtoPhot1}
	M. Hafezi, Z. Kim, S.L. Rolston, L.A. Orozco, B.L. Lev, J.M. Taylor, Phys. Rev. A \textbf{85}, 020302(R) (2012). 

	\bibitem{MWtoPhot2}
	L. A. Williamson, Y.H. Chen, and J. J. Longdell, Phys. Rev. Lett. \textbf{113}, 203601 (2014).

	\bibitem{MOProp1}
	A. S. S{\o}rensen, C. H. van der Wal, L. I. Childress, and M. D. Lukin, Phys. Rev. Lett. {\bf 92}, 063601 (2004).	
	
	\bibitem{NV1}
	D. Marcos, M. Wubs, J.M. Taylor, R. Aguado, M. D. Lukin, A.S. S{\o}rensen, Phys. Rev. Lett. {\bf 105}, 210501 (2010).
		
	\bibitem{NV3}
	A. Sipahigil, M.L. Goldman, E. Togan, Y. Chu, M. Markham, D.J. Twitchen, A.S. Zibrov, A. Kubanek, and M.D. Lukin,  Phys. Rev. Lett. {\bf 108}, 143601 (2012).
	
	\bibitem{zollerPolar}
	A. Andr\'{e} \textit{et al.}, Nat. Phys. \textbf{2}, 636-642 (2006).
	
	\bibitem{decoherenceQuasi}
	L. R. Testardi, Phys. Rev. B \textbf{4}, 2189 (1971).
		
	\bibitem{transformLimited}
	A. V. Kuhlmann et al Nat. Comm. {\bf 6}, 8204 (2015).
	
	\bibitem{lodahl}
	M. Arcari \textit{et al.}, Phys. Rev. Lett. {\bf113}, 093603 (2014).
	
	\bibitem{DasPRL17}
	S. Das, V. E. Elfving, S. Faez, and A. S. S{\o}rensen, Phys. Rev. Lett. {\bf 118}, 140501 (2017).
	
	\bibitem{imamoglu}
	Y. Tsuchimoto, P. Knuppel, A. Delteil, Z. Sun, M. Kroner, and A. Imamo{\u{g}}lu, Phys. Rev. B {\bf 96}, 165312 (2017).
	
	\bibitem{Dicke}
	R.H. Dicke, Phys. Rev. \textbf{93} 1, 99-110 (1954).
		
	\bibitem{haroche}
	M. Gross, S. Haroche, Phys. Rep. \textbf{93}, 5 (1982).
		
	\bibitem{kaiser}
	W. Guerin, M. O. Ara\'{u}jo, and R. Kaiser,	Phys. Rev. Lett. \textbf{116}, 083601 (2016).
	
	\bibitem{Solano17}
	P. Solano, P. Barberis-Blostein, F. K. Fatemi, L. A. Orozco, and S. L. Rolston, Nat. Comm. {\bf 8}, 1857 (2017).	
	
	\bibitem{goban}
	A. Goban, C.L. Hung, J.D. Hood, S.P. Yu, J.A. Muniz, O. Painter, and H.J. Kimble, Phys. Rev. Lett. \textbf{115}, 063601 (2015).	
	
	\bibitem{laurat2}
	N.V. Corzo, B. Gouraud, A. Chandra, A. Goban, A.S. Sheremet, D.V. Kupriyanov, and J. Laurat, Phys. Rev. Lett. \textbf{117}, 133603 (2016).
		
	\bibitem{coherencecpb}
	D. Vion \textit{et al.}, Science \textbf{296}, 5569, 886-889 (2002). 
	
	\bibitem{nakamura}
	Y. Nakamura, Yu. A. Pashkin and J. S. Tsai, Nature \textbf{398},  786-788 (1999).
		
	\bibitem{devoretReview}
	M. H. Devoret, R. J. Schoelkopf, Science \textbf{339}, 1169-1174 (2013).

	\bibitem{singtrip1}
	N. Samkharadze \textit{et al.}, Science  \textbf{359},  1123-1127 (2018).
	
	\bibitem{singtrip2}
	A. Stockklauser \textit{et al.}, Phys. Rev. X \textbf{7}, 011030 (2017).
	
	\bibitem{singtrip3}
	X. Mi, J. V. Cady, D. M. Zajac, P. W. Deelman, and J. R. Petta, Science \textbf{355}, 156-158  (2017).
	
	\bibitem{stark}
	P.W. Fry, I.E. Itskevich, D. J. Mowbray, M.S. Skolnick, J.J. Finley, J.A. Barker, E.P. OReilly, L.R. Wilson, I.A. Larkin \textit{et al.}, Phys. Rev. Lett. {\bf 84}, 733 (2000).
	
	\bibitem{Reiter12}
	F. Reiter and A. S. S{\o}rensen, Phys. Rev. A \textbf{85}, 032111 (2012).
	
	\bibitem{Das17}
	S. Das, V. E. Elfving, F. Reiter, A. S. S{\o}rensen, Phys. Rev. A \textbf{97}, 043837 (2018)
	
	\bibitem{singlephotonChang}	
	D. E. Chang, A. S. S{\o}rensen, E. A. Demler \& M D. Lukin, Nat. Phys. \textbf{3}, 807–812 (2007).
	
	\bibitem{singlesided}
	D. Witthaut and A. S. S{\o}rensen, New J. Phys \textbf{12}, 043052 (2010).
		
	\bibitem{Can15}
	T. Caneva \textit{et al.}, New J. Phys. \textbf{17}, 113001 (2015).
			
	\bibitem{tommaso}
	T. Pregnolato \textit{et al.}, Preprint at https://arxiv.org/abs/1907.01426  (2019).
	
	\bibitem{supp}
	See Supplementary Information for details of the two-QD transducer analytics and numerics, for unequal QD decay rates.
		
	\bibitem{QDpurcell1}
	V. S. C. Manga Rao and S. Hughes, Phys. Rev. B \textbf{75}, 205437 (2007).
	
	\bibitem{QDpurcell2}
	T.B. Hoang \textit{et al.}, App. Phys. Lett. \textbf{100}, 061122 (2012).
	
	\bibitem{Solano_arXiv}
	P. Solano \textit{et al.}, arXiv:1704.08741 (2017).
	
	\bibitem{gaas}
	D. E. Aspnes,  S. M. Kelso,  R. A. Logan,  and R. Bhat, J. Appl. Phys \textbf{60}, 754-767 (1986).
	
	\bibitem{aluminum}
	A. D. Raki\'{c},  Appl. Opt. \textbf{34}, 4755-4767 (1995).
		
	\bibitem{MachZender}
	C. Cabrillo, J. I. Cirac, P. Garc\"{i}a-Fern\'{a}ndez, and P. Zoller,	Phys. Rev. A {\bf 59}, 1025 (1999).
	
\end{thebibliography}
\end{document}